\documentstyle[prl,aps]{revtex}
\input epsf

\tighten

\begin{document}

%\rline{CWRU-P29-98}

\newcommand{\Bd}{{\dot B}}
\newcommand{\Cd}{{\dot C}}
\newcommand{\fd}{{\dot f}}
\newcommand{\hd}{{\dot h}}
\newcommand{\ep}{\epsilon}
\newcommand{\vp}{\varphi}
\newcommand{\al}{\alpha}
\newcommand{\be}{\begin{equation}}
\newcommand{\ee}{\end{equation}}
\newcommand{\bea}{\begin{eqnarray}}
\newcommand{\eea}{\end{eqnarray}}
\def\gapp{\mathrel{\raise.3ex\hbox{$>$}\mkern-14mu
              \lower0.6ex\hbox{$\sim$}}}
\def\gsim{\gapp}
\def\lapp{\mathrel{\raise.3ex\hbox{$<$}\mkern-14mu
              \lower0.6ex\hbox{$\sim$}}}
\def\lsim{\lapp}
\newcommand{\PSbox}[3]{\mbox{\rule{0in}{#3}\includegraphics{#1}\hspace{#2}}}
\def\Tr{\mathop{\rm Tr}\nolimits}
\def\su#1{{\rm SU}(#1)}

\title{Interaction of Magnetic Monopoles and Domain Walls}

\author{
Levon Pogosian
and
Tanmay Vachaspati}
\address
{Department of Physics,
Case Western Reserve University,
10900 Euclid Avenue,
Cleveland, OH 44106-7079, USA.}

\wideabs{
%\twocolumn[
\maketitle

\begin{abstract}
\widetext
We study the interaction of magnetic monopoles and 
domain walls in a model with SU(5)$\times Z_2$ symmetry
by numerically evolving the field equations.
We find that the monopoles unwind and dissipate their
magnetic energy on collision with domain walls within 
which the full SU(5) symmetry is restored.  
\end{abstract}
\pacs{}
}
%]

\narrowtext

The interactions of topological defects can have a profound
effect on the outcome of phase transitions. The scaling
of a network of domain walls and strings, and a distribution
of magnetic monopoles, crucially depends on how the defects
interact among themselves and with each other. Thus far
attention has focussed on the interactions of walls with
walls, strings with strings, and monopoles with monopoles.
The cosmological importance of the interactions of walls and 
monopoles was highlighted in Ref. \cite{DvaLiuVac97} and it
is this problem that we study in the present paper.

Earlier work on the interaction of solitons and domain
walls (phase boundaries) has been carried out in the 
following contexts: (i) mutual interaction of domain walls 
\cite{She86}, (ii) He$^3$ A-B phase boundaries and vortices 
\cite{TreKutetal}, (iii) Skyrmions and domain walls \cite{piette}, 
and (iv) global monopoles and embedded domain walls in an O(3)
linear $\sigma$ model \cite{Ale99}. Here we will
numerically study the interaction of gauged SU(5) monopoles 
with a Z$_2$ domain wall. This is quite distinct from the 
earlier work since it looks at magnetic monopoles which
necessarily include gauge fields. It is also the most
relevant problem for the cosmological consequences of
Grand Unified theories \cite{DvaLiuVac97}. 

The SU(5) model we consider is given by the Lagrangian:
\begin{equation}
L = -{1\over 4} X^a_{\mu\nu}X^{a\mu\nu} + 
     {1\over 2} (D_\mu\Phi^a )^2 - V(\Phi ) \ ,
\label{lagrangian}
\end{equation}
where $\Phi$ is an SU(5) adjoint scalar field, 
$X^a_{\mu\nu}$ ($a=1,...,24$) are the gauge
field strengths and the covariant derivative is defined by:
\begin{equation}
D_\mu \Phi^a = \partial_\mu \Phi^a - i e [X_\mu ,\Phi]^a  
\label{covariantderivative}
\end{equation}
and the group generators are normalized by 
${\rm Tr}(T_a T_b)=\delta_{ab}/2$.
The potential $V(\Phi )$ is the most general quartic potential
but we exclude the cubic term in $\Phi$ so as to obtain the
extra Z$_2$ symmetry under $\Phi \rightarrow -\Phi$:
\begin{equation}
V(\Phi ) = -m^2 {\rm Tr}\Phi^2 + 
            h ({\rm Tr} \Phi^2 )^2 +
               \lambda {\rm Tr} \Phi^4 \ .
\label{potential}
\end{equation}
The parameters of the potential are chosen so that 
$\langle \Phi \rangle = \eta {\rm diag}(2,-3,2,2,-3)/(2\sqrt{15})$ with 
$\eta =m/\sqrt{\lambda '}$ and $\lambda ' = h+7\lambda/30$.
With this vacuum expectation value,
the SU(5) symmetry is spontaneously broken to 
SU(3)$\times$SU(2)$\times$U(1). The desired constraints on 
the parameters are: $\lambda , \lambda ' > 0$.

The magnetic monopoles in this model were discussed by
Dokos and Tomaras \cite{DokTom} except that also
included the effects of a scalar field in the fundamental
representation of SU(5). Here we do not have such a field.
Yet the basic construction of \cite{DokTom} goes through
and the fundamental monopole is essentially an SU(2) monopole
embedded in the full theory. 
The monopole solution has the following form:
\begin{equation}
\Phi_M \equiv \sum_{a=1}^{3} \Phi^a T^a + \Phi^4 T^4 +
\Phi^5 T^5 \ ,
\label{monopolesolution}
\end{equation}
where the subscript $M$ denotes the monopole field configuration,
$$
T^a = \frac{1}{2}{\rm diag}(\sigma^a,0,0,0)\ , \ \ 
T^4 = \frac{1}{2\sqrt{3}}(0,0,1,1,-2) \ ,
$$
$$
T^5 = \frac{1}{2\sqrt{15}}(-3,-3,2,2,2) \ ,
$$
$\sigma^a$ being the Pauli spin matrices, 
\begin{equation}
\Phi^a = P(r) x^a \ , \
\Phi^4 = M(r) \ , \
\Phi^5 = N(r) \ ,
\end{equation}
where $r=\sqrt{x^2+y^2+z^2}$ is the spherical radial coordinate.
The ansatz for the gauge fields for the monopole is: 
$$
W^a_i = \epsilon^a_{~ij} 
\frac{x^j}{er^2}(1-K(r))  \ , \ 
(a=1,2,3) \ ,
$$
\begin{equation}
W^b_i = 0, \ , \ \ (b \ne 1,2,3).
\label{gaugeansatz}
\end{equation}

In the case when the 
potential vanishes (the Bogomolnyi-Prasad-Sommerfield (BPS) 
case \cite{Bog,PraSom}), the exact solution is known \cite{Mec99}:
\begin{equation}
P(r) = \frac{1}{er^2}(\frac{Cr}{\tanh(Cr)}-1) \ , \
K(r) = \frac{Cr}{\sinh(Cr)} \ ,
\label{bpsphiprofile}
\end{equation}
\begin{equation}
M(r) = \frac{2}{\sqrt{3}}\frac{C}{e} \ , \ \
N(r) = \sqrt{\frac{1}{15}}\frac{C}{e} \ .
\end{equation}

In the non-BPS case, the profile functions $P(r)$, $K(r)$, $M(r)$ and
$N(r)$ need to be found numerically. We find them by 
using a relaxation procedure with
the BPS solution serving as the initial guess.

Depending on the parameters in the potential, it is possible
to have different stable domain wall solutions. The domain
wall across which $\Phi \rightarrow - \Phi$ is stable provided
\cite{DvaLiuVac97}, 
\begin{equation}
-{3 \over {20}} > {h \over {\lambda}} > -{7 \over {30}}  \ .
\label{hoverlambda}
\end{equation}
At the center of this wall, $\Phi$ must necessarily vanish
and so the full SU(5) symmetry is restored at the center of
this wall. Certain components of $\Phi$ do not vanish at
the center of the domain wall solutions in this model for
other values of parameters. In these walls,
only a subgroup of the full SU(5) symmetry is restored in the
center. We will only study the interaction of monopoles with 
walls in which $\Phi=0$ at the center in this paper. The 
interactions of other types of walls and monopoles will be
discussed separately.

The solution for the domain wall located in the xy-plane is
\begin{equation}
\Phi_{DW} = {\eta \over {2\sqrt{15}}} \tanh(\sigma z) (2,-3,2,2,-3) \ ,
\end{equation}
where $\sigma=\eta \sqrt{\lambda '/2}$.

When the monopole and the domain wall are very far from each other, 
the joint field configuration is given by the product ansatz:
\begin{equation}
\Phi = \tanh (\gamma \sigma (z-z_0)) \Phi_M \ ,
\label{initialfield}
\end{equation}
where $v$ is the velocity of the domain wall in the negative z-direction, 
$\gamma=1/\sqrt{1-v^2}$ is the Lorentz factor and $z_0$ is the 
position of the wall. Here $\Phi_M$ denotes the monopole
solution in eq. (\ref{monopolesolution}).
The gauge fields are unaffected by the presence of the wall
and are still given by eq. (\ref{gaugeansatz}).
In addition, the time derivative of the scalar field is also
given by the product ansatze:
\begin{equation}
\dot{\Phi} = \gamma \sigma v \ $sech$^2(\gamma \sigma (z-z_0)) \Phi_M \ .
\label{initialdotfield}
\end{equation}

Eqs. (\ref{initialfield}) and (\ref{initialfield}) specify the 
initial ($t=0$) conditions for the
scalar field for a  wall approaching a monopole with velocity $v$. 
The initial scalar and gauge field profile functions
$P$, $M$, $N$ and $K$ (in the non-BPS case) are found by numerical 
relaxation. The field dynamics is 
described by the equations of motion following from the Lagrangian 
in (\ref{lagrangian}). At first sight, there are 24 components of 
$\Phi$ and 96 components of the gauge fields that need to be evolved. 
However, it is not hard to check that all the dynamics occurs in an
SU(2) subgroup of the original SU(5). This then reduces the
dynamical fields to a triplet of SU(2) and two other fields (i.e. a total of 5 
scalar fields) and 3$\times$4=12 gauge field components. Choosing the
temporal gauge ($W^a_0 =0$) reduces the number of 
gauge field components to 9.

Further reduction of the problem occurs since the initial
conditions are axially symmetric and the evolution equations
preserve this symmetry. The angular dependence in cylindrical
coordinates can easily be imposed on the scalar field. For the
gauge fields it can be extracted by using the
fact that the covariant derivatives of the scalar field must
vanish at large distances from the monopole. This then leads
to the following ansatz for the 5 scalar and 9 gauge fields:
\begin{eqnarray}
&\Phi_1 = f_1~ x \ , \ \  \Phi_2 = f_1~ y\ , 
\ \ \Phi_3 = f_2~ z \nonumber \\ 
&\Phi_4 = f_3 \ , \ \ \Phi_5 = f_4 \nonumber \\
&W^1_x = f_5~ xy \ , W^1_y = f_5 ~ y^2 - f_6 \ , W^1_z = f_7~  y
\nonumber \\
&W^2_x = -f_5~ x^2+f_6 \ , W^2_y = -f_5~  xy \ , W^2_z = -f_7 ~ x
\nonumber \\
&W^3_x = - f_8~ y \ , W^3_y = f_8~ x \ , W^3_z = 0 \ , \nonumber 
\end{eqnarray}
where the $f_i$ ($i=1,...,8$) are functions only of $t$, 
$\rho=\sqrt{x^2+y^2}$ and $z$. We have explicitly checked that
this ansatz is preserved by the evolution equations.
So now the problem is reduced to one in 8 real functions of
time and two spatial coordinates.

An attempt to numerically solve the 8 equations of motion 
directly in cylindrical coordinates failed due to numerical
instabilities that developed within the time scale of the 
simulation. An analysis showed
that the problem was due to large numerical errors in evaluating
the derivatives in cylindrical coordinates. This
shortcoming of using cylindrical (and spherical) coordinates
in numerical work is well-recognized and the authors
of \cite{numericalpaper} have proposed a solution that we
have successfully implemented. The idea is to solve the problem,
not in two spatial dimensions like the $\rho$z-plane, but to solve
it in a thin three dimensional slab whose central slice is
taken to lie in the xz-plane and
with only 3 lattice spacings along the y direction. Then Cartesian 
coordinates can be used to solve the equations of motion in the 
$y=0$ plane, thus minimizing numerical errors. On the $y\ne 0$ 
lattice sites the fields are evaluated by using the axial symmetry
of the problem. This scheme improved the numerical stability
of our staggered leapfrog code dramatically and allowed us to
observe the monopole and wall for a sufficiently long duration
without the development of numerical instabilities.

We have evolved the initial wall and monopole configuration
with several velocities. The numerical results of the simulation with 
$v=0.8$, $h=-\lambda/5$, $\lambda=0.5$ and $\eta=1$
are given in the figures and clearly show that the energy of the monopole 
dissipates after the passage of the wall ($m=\eta\sqrt{\lambda '}$ and $e$ 
can be scaled out of the problem). The final snapshot shows that
the energy in the scalar field is located entirely on the wall and 
the magnetic energy is along and behind the wall. 

When the domain wall moves close to the monopole the latter is pulled 
toward the wall. This signals
the presence of an attractive force between the two defects. 
Such a force is expected from energy considerations \cite{DvaLiuVac97}
and has been observed in the O(3)
linear $\sigma$ model studied in \cite{Ale99}.

We have estimated the time it takes for the monopole to be destroyed
as a function of different wall velocities. The topological winding 
density of the monopole is a delta function in space, while the topological
winding is a discrete number and so these quantities cannot be used to 
estimate the dissolution time. Instead, the magnetic energy density provides
us with a continuously varying quantity that we can track in the 
simulations. (The scalar energy is not suitable since both the wall 
and the monopole contribute.) We have chosen a cylindrical volume 
around the monopole with axis along the z-axis and computed the 
magnetic energy within this volume as a function of time. This gives
us the rate at which the magnetic energy escapes the cylindrical volume 
surrounding the monopole. There are two stages in the dissolution
process - first the monopole gets absorbed by the wall and then the
magnetic energy starts spreading in the direction along the wall. 
The time taken for the monopole to be absorbed by the wall is
measured from when the cores overlap (corresponding to a sharp drop
in the magnetic energy and a sharp rise in the electric energy) 
to when the magnetic energy starts moving together with the wall
in the z-direction but spreading along the wall. In our simulations, 
we observe 
that the absorption time is approximately equal to the width of the wall 
in the monopole rest frame, that is $t_{abs} \sim (\gamma \sigma)^{-1}$.
During the second stage the remaining magnetic energy is confined to
the wall and is escaping the cylinder as it travels along the wall.
In the rest frame of the wall the duration of the second step is
independent of the wall velocity but takes longer in the initial 
rest frame of the monopole by a factor that is well-accounted for by 
time dilation. 

At high wall velocities, Lorentz contraction results in the wall
appearing to be much ``thinner'' than the monopole. We have tried 
velocities as high as $v=0.99$ and have not observed any qualitative
alterations from the dynamics at lower velocites except for the obvious
time dilation of the propagation of the excitaions along the wall. 

The above results have not changed as we varied the parameters in the 
potential within the range in which a $\Phi \rightarrow - \Phi$ domain 
wall solution is stable. To better understand this 
independence of the dynamics on the parameters let us examine
the parameter space itself and the characteristic length scales that
are involved. The stability of the domain wall is determined 
by the ratio $h/\lambda$. The value of $\eta$ sets the 
fundamental time and length scales of the simulation and can be used
to adjust the lattice grid spacing to the ``widths'' of the defects.
There are three relevant length scales in the problem: 
the sizes of the scalar and vector cores of the monopole, $r_S$ and $r_V$, 
and the thickness of the domain wall, $r_{DW}$. These are
the sizes of the regions in which the fields deviate
significantly from their asymptotic values. 
We can estimate $r_S$, $r_V$ and $r_{DW}$ numerically by 
finding the distance at which the corresponding fields become an exponent 
closer to their respective asymptotic values. From dimensional agruments $r_S$ 
and $r_V$ are approximately inversely proportional to the masses of 
the scalar and vector bosons making up the monopole, 
$m_S = \eta \sqrt{\lambda'}$ and $m_V = \sqrt{5/12} \, \ e \eta$. The
thickness of the domain wall is given by 
$\sigma^{-1}=\eta^{-1}\sqrt{2/\lambda '}$. 
The interaction between the monopole and the wall depends 
on the relative values of $r_S$ and $r_{DW}$. The
dimensional arguments give $r_S \sim r_{DW}$ while numerically we
find $r_S < r_{DW}$ for all values of $h / \lambda$
that lead to stable domain walls (see eq. (\ref{hoverlambda})) and $\lambda 
\in [0.01,100]$.
For sufficiently high values of $\lambda$ it is possible to have
the vector width of the monopole to be larger than the domain wall
width, i.e.
$r_V>r_{DW}>r_S$. Our simulations show that
the walls still sweep the monopoles. 
We will study the many different kinds of domain walls that can occur 
in the model and their interactions with monopoles separately. 

\begin{figure}[tbp]
\vskip 0.1 truein
\epsfxsize = 0.8 \hsize  \epsfbox{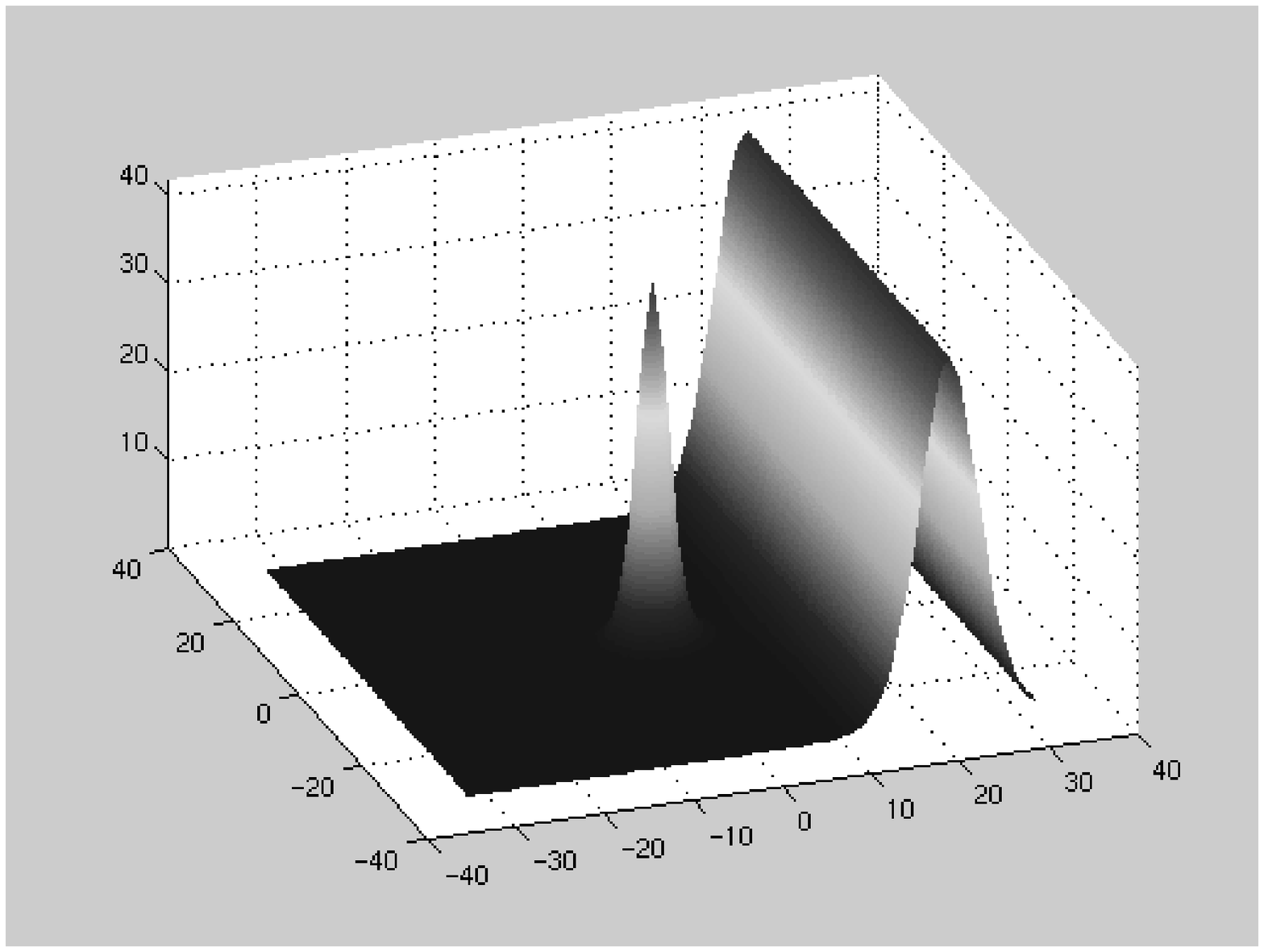}
\vskip 0.1 truein
\epsfxsize = 0.8 \hsize  \epsfbox{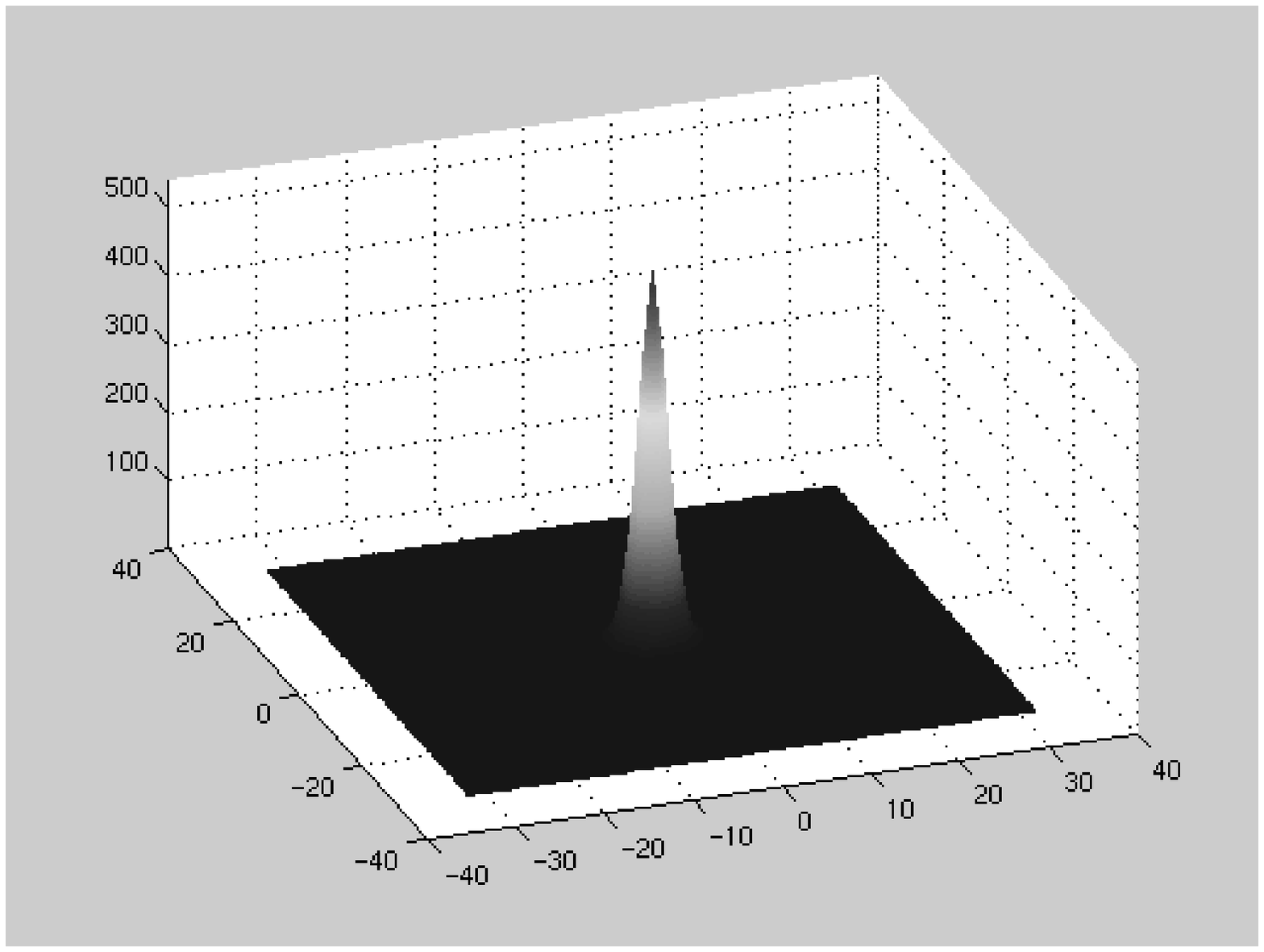}
\vskip 0.2 truein
\caption{\label{figure1}
The first panel shows the potential energy density in the 
xz-plane for the 
magnetic monopole and domain wall where $h=-\lambda/5$, 
$\lambda=0.5$ and the wall 
velocity is $0.8c$.
The second panel shows the corresponding magnetic energy density 
(proportional to ${B^a}_i^2$).
}
\end{figure}

\vfill

\begin{figure}[tbp]
\epsfxsize = 0.8 \hsize  \epsfbox{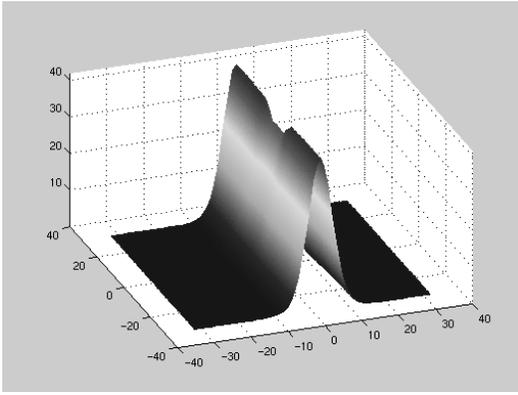}
\vskip 0.1 truein
\epsfxsize = 0.8 \hsize  \epsfbox{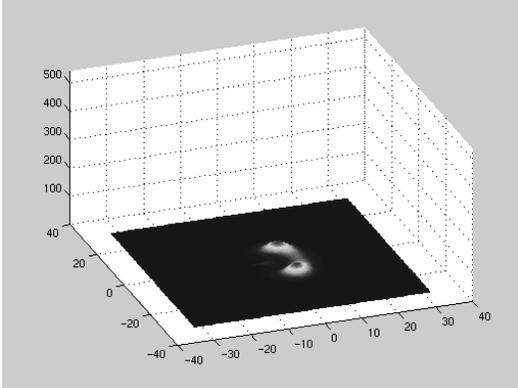}
\caption{\label{figure2}
As in Fig. \ref{figure1} at an intermediate time step.
}
\end{figure}

\vfill

\begin{figure}[tbp]
\epsfxsize = 0.8 \hsize  \epsfbox{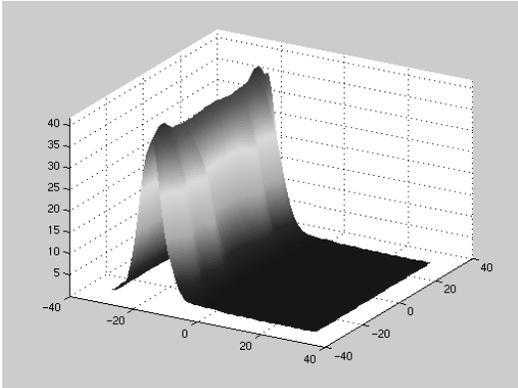}
\vskip 0.1 truein
\epsfxsize = 0.8 \hsize  \epsfbox{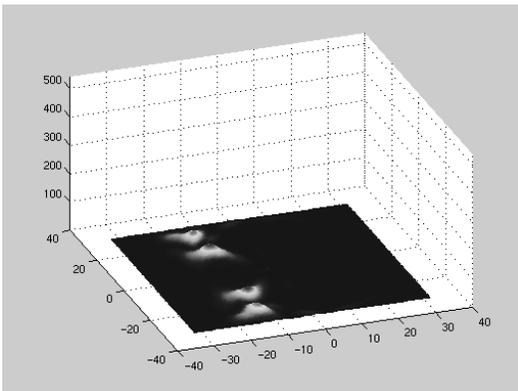}
\caption{\label{figure3}
As in Fig. \ref{figure1} at the final time step.
}
\end{figure}

As is clear from Fig. \ref{figure3}, the final state of the wall
is different from that of the initial wall. The destruction of
the monopole has left a residue of scalar and magnetic
excitations on the domain wall that are propagating
along and behind the wall.

The dissolution of magnetic monopoles by domain walls implies that
the number density of magnetic monopoles will fall off faster than
if there were no domain walls. The cosmology of such a system of walls
and monopoles has been discussed in \cite{DvaLiuVac97} where it was
argued that such interactions might resolve the cosmological
monopole over-abundance problem. Similar interactions between strings
and domain walls would affect the cosmological implications of
cosmic strings. The numerical techniques presented 
here can also be used to study the interactions of walls and (global) 
monopoles or vortices in other systems.

%\acknowledgements
We would like to thank Mark Meckes for providing the
SU(5) BPS monopole solution, Ken Olum for numerical
advice and especially Paul Saffin for bringing 
Ref. \cite{numericalpaper} to our attention. We thank the
the National Energy Research Scientific Computing Center
(NERSC) for the use of their J90 cluster on which some of the
simulations were run.
This work was supported by the Department of Energy (DOE).

\end{document}